\documentclass[twocolumn,prb,showpacs,preprintnumbers,amsmath,amssymb]{revtex4}
\usepackage{graphicx}
\usepackage{dcolumn}
\usepackage{bm}

\begin{document}
\title{Unmasking the nodal quasiparticle dynamics in cuprate superconductors \\ using low-energy photoemission}
\author{T. Yamasaki,$^1$ K. Yamazaki,$^1$ A. Ino,$^1$ M. Arita,$^2$ H. Namatame,$^2$ M.~Taniguchi,$^{1,2}$ \\ A. Fujimori,$^3$ Z.-X. Shen,$^4$ M. Ishikado,$^5$ and  S. Uchida$^5$}

\affiliation{$^1$Graduate School of Science, Hiroshima University, Higashi-Hiroshima 739-8526, Japan}
\affiliation{$^2$Hiroshima Synchrotron Radiation Center (HSRC), Hiroshima University, Higashi-Hiroshima 739-8526, Japan}
\affiliation{$^3$Department of Complexity Science and Engineering, University of Tokyo, Kashiwa 277-8561, Japan}
\affiliation{$^4$Department of Physics, Applied Physics and Stanford Synchrotron Radiation Laboratory, Stanford University, Stanford, CA 94305, USA}
\affiliation{$^5$Department of Physics, University of Tokyo, Tokyo 113-0033, Japan}

\begin{abstract}
Nodal quasiparticles of Bi$_2$Sr$_2$CaCu$_2$O$_{8+\delta}$ have been studied by angle-resolved photoemission spectroscopy with high momentum and energy resolution.  Low-energy tunable photons have enabled us to resolve a small nodal bilayer splitting, unmasking the intrinsic single-particle scattering rate.  The nodal scattering rate is abruptly suppressed upon the superconducting transition, and shows a linear energy dependence at low energies, indicating the nontrivial effect of elastic scatterings on the quasiparticles.  With increasing energy, the antibonding-band scattering rate becomes higher than the bonding one.  The observations imply the character of the scatterers dominant at low energies. 
\end{abstract} 
\pacs{74.72.Hs, 74.25.Jb, 79.60.-i}
\maketitle

Thermodynamics properties of a superconductor both in the zero-field and vortex-mixed states are governed by excitations of low-energy quasiparticles.  They are particularly important for a $d$-wave superconductor, where the gap has a node and therefore the excitations start from the zero energy.  Nevertheless, probing the nodal scattering rate is often difficult due to the masking by the superconductivity itself.  To date, microwave and optics experiments have indicated the presence of well-defined quasiparticles in high-$T_c$ cuprate superconductors.\cite{Bonn1,Krishana,Hosseini,Puchkov,Takeya}  However, the transport properties are intricately integrated and require some assumptive models, e.g., the extended Drude model, to extract individual quasiparticle properties.  Especially in bilayer cuprates, the band is split due to the interaction between two proximate CuO$_2$ planes,\cite{Feng,Chuang} making the modeling more difficult.

On the other hand, angle-resolved photoemission spectroscopy (ARPES) directly probes single-quasiparticle excitations resolved in energy-momentum space.  The quasiparticle scattering rate and dispersion are simultaneously determined from the spectral-peak width and position, respectively.  Hence, serious ARPES studies have been attempted for the node.\cite{ARPES,Review,Lanzara,Bogdanov,VallaNew} Recently, it has been shown that the bilayer splitting remains finite even in the nodal direction, but direct resolution of individual quasiparticle properties has been difficult due to the proximity of two bilayer states.\cite{Kordyuk2}  Based on the analysis of circular-polarization-dependent spectra, the narrower spectral peak width for the antibonding band has been reported as an implication of dominant magnetic scattering.\cite{Borisenko}  On the other hand, in an even higher-resolution experiment using a 6-eV laser, only single band dispersion has been observed in the same nodal direction.\cite{Koralek}  This discrepancy casts uncertainties upon interpreting these previous data.

In the present paper, we show bilayer-resolved nodal scattering rates, directly obtained by using low-energy tunable photons ($h\nu = 7.57$ eV).  The high momentum resolution and the photon-energy tunability enabled us to visualize the sharp ($\Delta k = 0.0065$ {\AA}$^{-1}$) and complete image of the bilayer-split nodal spectral function.  Based on the detailed energy, temperature and bilayer-band dependences of the scattering rates, we argue that the low-energy quasiparticles are seriously affected by the elastic scattering process.  The difference in the bilayer-resolved scattering rates implies the spatial distribution of the scatterers dominant at low energies.

ARPES using low-energy excitation photons has an advantage in realizing high energy and momentum resolution.\cite{Koralek}  Because in-plane quasiparticle momenta $k_{\parallel}$ are magnified into large emission angles for low excitation energies, the momentum resolution improves appreciably for a given instrumental angular resolution.  In addition, the presumable increase in the photoelectron escape depth\cite{Seah} would minimize the effect of possible surface imperfection and contamination, and suppress the smearing in surface-normal momentum $k_{\perp}$, because the well-defined final state imposes a strict condition on photoelectron transition.  On the other hand, the photoelectron final state is no longer like a free electron for low-energy excitation.  Therefore, tuning the photon energy to the final state is critically important.  Indeed, we find that the photon-energy dependence is severe in low-energy ARPES.  In the present study, we optimized the photon energy to $h\nu =7.57$ eV so that both the bonding and antibonding bands are clearly observed at once.

The ARPES experiments were performed at a helical undulator beamline, BL-9 at Hiroshima Synchrotron Radiation Center, using circularly polarized light and a SCIENTA SES2002 analyzer.  The total energy resolution was set at 4 meV.  Samples are nearly optimally doped Bi$_2$Sr$_2$CaCu$_2$O$_{8+\delta}$ ($T_{c} = 86$ K), which were cleaved {\it in situ} under an ultrahigh vacuum, $< 1{\times}10^{-10}$ Torr.

\begin{figure}
    \includegraphics[width=8.6cm]{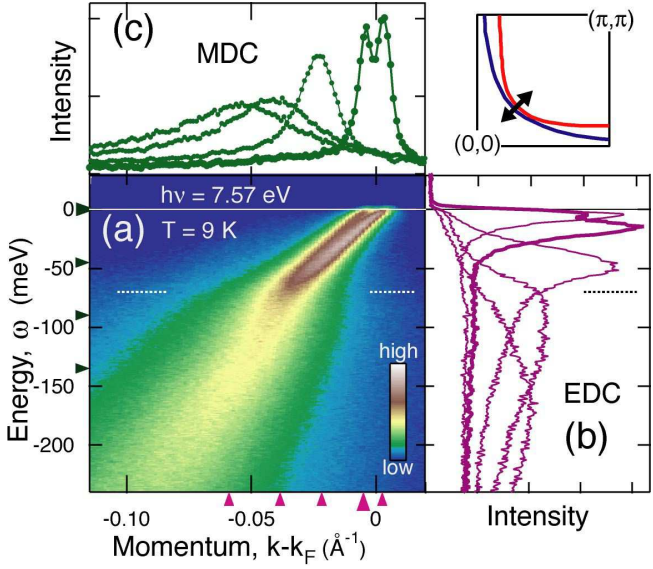} \vspace{-1.5pc}
    \caption{Overview of low-energy ARPES result at $h{\nu} = 7.57$ eV, taken in the nodal direction of superconducting Bi$_2$Sr$_2$CaCu$_2$O$_{8+\delta}$ at $T = 9$ K.  (a) Spectral-intensity map in the energy-momentum space.  (b) Energy distribution curves (EDCs) at the momenta denoted by purple triangles.  (c) Momentum distribution curves (MDCs) at the energies denoted by green triangles.  The intensity of the MDC at ${\omega} = 0$ is multiplied by 2.}
    \label{overview}
\end{figure}

\begin{figure}
    \vspace{1pc} \includegraphics[width=8.6cm]{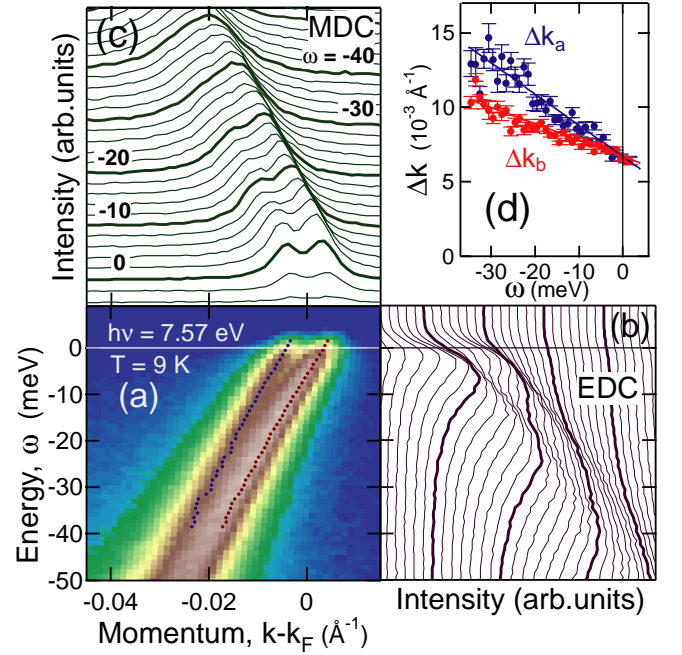} \vspace{-1.5pc}
    \caption{Enlarged view of Fig.~\ref{overview} around the Fermi-surface crossings.  (a) Spectral-intensity map.  Red and blue circles denote the peak positions of the MDCs. (b) EDCs at each 0.0012 {\AA}$^{-1}$.  (c) MDCs at each 2 meV.  (d) Bilayer-resolved scattering rates, determined from the momentum widths (FWHM), ${\Delta}k_b$ (red), and ${\Delta}k_a$ (blue), of the bonding and antibonding peaks, respectively.  Thin solid lines show the linear increasing rates, $\sim 0.11$ and $\sim 0.21$ \AA$^{-1}$ eV$^{-1}$, of ${\Delta}k_b$ and ${\Delta}k_a$, respectively.}
    \label{enlarged}
\end{figure}

Results at a low temperature of $T = 9$ K are shown in Figs.~\ref{overview} and \ref{enlarged}, which demonstrate the performance of the low-energy ARPES.  First, Fig.~\ref{overview} clearly shows that the nature of quasiparticle excitations dramatically changes at an energy of $| \omega | \sim70$ meV.\cite{Lanzara,Bogdanov}  The quasiparticle peak dramatically sharpens on crossing $| \omega | \sim 70$ meV towards the Fermi level, in good correlation with the abrupt deceleration of the quasiparticle group velocity $v_k = d \omega_k / dk$.  Second, as shown in Fig.~\ref{enlarged}, the peak becomes very sharp near ${\omega} = 0$, so that the quite small splitting of 0.0075 {\AA}$^{-1}$ is distinctly resolved in contrast to the laser-ARPES spectra.\cite{Koralek}  The quasiparticle momentum width of $\Delta k = 0.0065$ \AA$^{-1}$ at $\omega = 0$ is much sharper than the previous ARPES data collected at higher photon energies,\cite{ARPES,Bogdanov} and reveals that the nodal quasiparticles travel for a long period, $\geq 150$ \AA, without scattering.

Peaks of the doublet are identified as the bonding and antibonding bands of two proximate CuO$_{2}$ layers.\cite{Feng,Chuang}  Such bilayer splitting has been treated as negligible at the node, because no Cu 4$s$ component hybridizes with the band of $d_{x^2-y^2}$ symmetry.\cite{Andersen} However, a small nodal splitting is allowed by intra-bilayer $p$-$p$ transfer.\cite{Kordyuk2}  We have carefully confirmed that the splitting width, $k_b - k_a = 0.0075\pm0.001$ \AA$^{-1}$, is reproduced for several samples.  Multiplying $k_b - k_a$ by the Fermi velocity $v_{F} = 1.9$ eV{\AA} gives the nodal splitting energy, $\omega_a - \omega_b = 14 \pm1$ meV, which is smaller than the earlier result\cite{Kordyuk2} and approximately 16{\%} of the antinodal splitting energy.\cite{Feng,Chuang}

The high-resolution spectral image of the split dispersion directly provides us the intrinsic scattering rates of the bonding and antibonding bands.  Figure~\ref{enlarged}(d) shows the widths of the bilayer-split peaks, resolved by fitting each momentum distribution curve (MDC) with two independent Lorentzians.  The energy dependences of both the scattering rates appear to be dominated by a linear term over the higher order terms unlike for normal metal.  While the two scattering rates are identical at $\omega=0$, the antibonding-band scattering rate becomes higher than the bonding one with energy.  Also, in the energy distribution curves shown in Fig.~\ref{enlarged}(b), the low-energy-side antibonding peak is broader than the bonding one for $|\omega| \sim 30$ meV.  This result is a contrast to the interpretation of the circular-polarization-dependent spectra.\cite{Borisenko}

\begin{figure}
    \includegraphics[width=8.6cm]{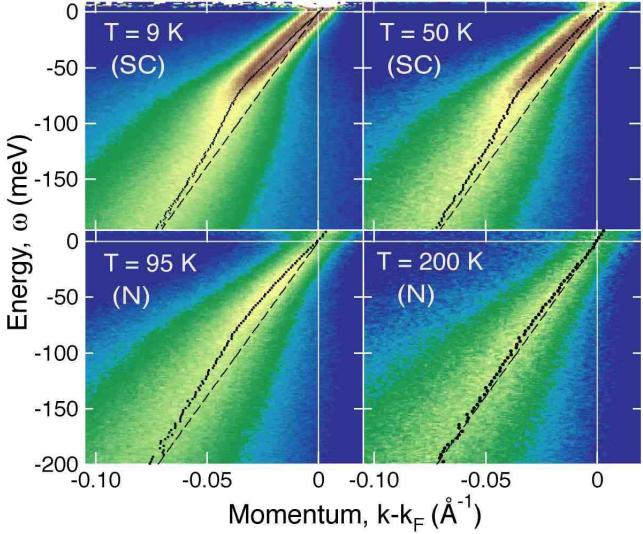} \vspace{-2pc}
    \caption{Single-particle spectral function in the superconducting ($T = 9$ and 50 K) and normal ($T = 95$ and 200 K) states, obtained by dividing the ARPES intensity by the Fermi-Dirac distribution function.  Dotted and thin dashed lines denote the weighted center of the bonding and antibonding peaks in the MDCs, and the hypothetical unrenormalized dispersion ${\epsilon_{k}^{0}}'= {v_0}'(k-k_{F})$  for ${v_0}' = 2.8$ eV \AA, respectively. } \vspace{-0.5pc}
    \label{spectral}
\end{figure}

Figure~\ref{spectral} shows the temperature dependence of the spectral function.  While the spectral feature is essentially unchanged within the superconducting state ($T = 9$ and 50 K), the quasiparticle peak near ${\omega} = 0$ broadens significantly in the normal state ($T = 95$ and 200 K).

\begin{figure}
    \includegraphics[width=8.6cm]{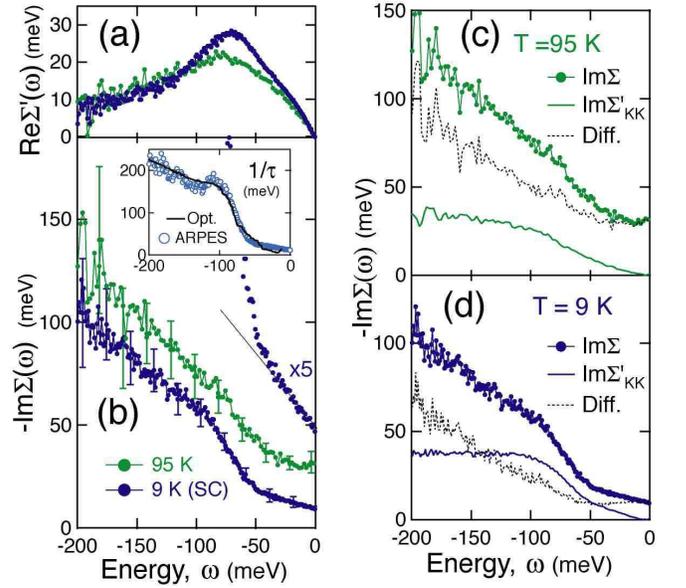} \vspace{-1.5pc}
    \caption{Self-energy $\Sigma(\omega)$ in the nodal direction, obtained by regarding the splitting parameters as constants.  (a) Real part of self-energy, determined from the quasiparticle dispersion by $\mathrm{Re}\Sigma'(\omega _{k})=\omega_k - {\epsilon_k^0}'$.  (b) Imaginary part of self-energy, determined from the quasiparticle momentum width $\Delta k$ (FWHM) by $\mathrm{Im}\Sigma(\omega) = -\frac{1}{2}v_{0} \Delta k$, where the scaling factor is $v_{0} = 2.8$ eV {\AA} (Ref.~\onlinecite{kdependence}).  Fivefold magnified view shows the $\omega$-linear fit (black line) of $\mathrm{Im}\Sigma(\omega)$ at $T = 9$ K.  Inset shows the inverse lifetimes, deduced from ARPES by $1/\tau (\omega) = v_k \Delta k$ (open circles) and from optical conductivity (line) (Ref.~\onlinecite{Puchkov}).  (c) and (d) Comparison between the width-derived $\mathrm{Im}\Sigma(\omega)$ (filled circles) and the Kramers-Kr\"{o}nig transformation $\mathrm{Im}\Sigma_{\mathrm{KK}}'(\omega)$ (solid lines) of the dispersion-derived $\mathrm{Re}\Sigma'(\omega)$ for $T = 95$ and 9 K.  Black dotted lines denote the difference, $\mathrm{Im}\Sigma(\omega) - \mathrm{Im}\Sigma_\mathrm{KK}'(\omega)$.} 
    \label{selfenergies}
\end{figure}

In order to extract the wide-range temperature and energy dependences of the scattering rate, we have performed the fitting analysis of all the MDCs, regarding splitting parameters as constants, i.e., relative peak position $k_b - k_a = 0.0075$ \AA$^{-1}$, width $\Delta k_a / \Delta k_b = 1$, and intensity $I_a / I_b = 0.78$, and obtained the momentum width and dispersion, which are averaged between the bonding and antibonding quasiparticles.  Consequently, the quasiparticle renormalization energy and scattering rate have been deduced as shown in Figs.~\ref{selfenergies}(a) and \ref{selfenergies}(b).  The real part of self-energy, $\mathrm{Re}\Sigma'(\omega)$, has been deduced from the dispersion deviation from the straight line shown in Fig.~\ref{spectral}.  The imaginary part of self-energy, $\mathrm{Im}\Sigma(\omega)$, has directly been determined from the momentum width $\Delta k$, i.e.\ the inverse scattering length, using the constant scaling factor of unrenormalized velocity $v_0 = 2.8$ eV \AA.\cite{kdependence}  The scattering rate, $\mathrm{Im}\Sigma(\omega)$, shows a clear steplike feature at $|\omega| \sim 70$ meV due to the coupling with optical phonon modes, and follows the linear $\omega$-dependence quite well at low energies in agreement with the bilayer-resolved data in Fig.~\ref{enlarged}(d).

\begin{figure}
    \vspace{2pc} \includegraphics[width=8.6cm]{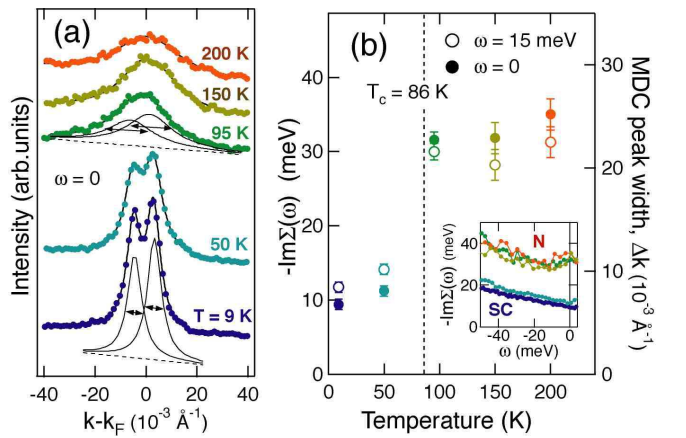} \vspace{-1.5pc}
    \caption{(a) Temperature dependence of the MDC at $\omega = 0$.  (b) Temperature dependence of the scattering rate, $-\mathrm{Im}\Sigma(\omega)=\frac{1}{2}v_0 \Delta k$, determined from the momentum width $\Delta k$ (FWHM) at $\omega = 0$ (filled circles) and 15 meV (open circles).  Inset shows the scattering rate as a function of energy for the superconducting (SC) and normal (N) states.} 
    \label{temperature}
\end{figure}

The consistency between the width-derived $\mathrm{Im}\Sigma(\omega)$ and the Kramers-Kr\"{o}nig transformation $\mathrm{Im}\Sigma_{\mathrm{KK}}'(\omega)$ of the dispersion-derived $\mathrm{Re}\Sigma'(\omega)$ is shown in Figs.~\ref{selfenergies}(c) and \ref{selfenergies}(d).  The step features of $\mathrm{Im}\Sigma(\omega)$ and $\mathrm{Im}\Sigma_{\mathrm{KK}}'(\omega)$ are consistent in the position $50 \leq |\omega| \leq 90$ meV and in the height $\sim35$ meV.  The bump at $|\omega|\sim90$ meV is cancelled out in the difference at all the temperatures.  The residual component, $\mathrm{Im}\Sigma(\omega) - \mathrm{Im}\Sigma_\mathrm{KK}'(\omega)$, is ascribed to the electron-electron scattering.  Upon taking the difference from a straight line as $\mathrm{Re}\Sigma'(\omega)$, the renormalization in high-energy scale, $|\omega| \gtrsim 100$ meV, is excluded, while that at $\sim 70$ meV is taken into account.  On the other hand, all kinds of scattering contribute to  $\mathrm{Im}\Sigma(\omega)$.  The inverse scattering time, $1/\tau (\omega)$, is also deduced from the momentum width multiplied by the quasiparticle group velocity, $1/\tau (\omega) = v_k \Delta k$.  The inset of Fig.~\ref{selfenergies}(b) shows the nodal single-particle scattering rate determined by ARPES and the transport scattering rate derived from the optical conductivity.\cite{Puchkov}  The overall similarity indicates that the extrinsic spectral broadening is minimal in the present ARPES study.

Figure~\ref{temperature} shows the temperature dependence of the MDC and scattering rate at ${\omega}$ = 0.  We have found that the momentum width abruptly drops by $\sim0.01$ \AA$^{-1}$ upon the superconducting transition ($T_c = 86$ K), indicating that the large portion of the nodal scattering rate, 60$-$70 {\%}, is suppressed in the superconducting state.  This discontinuity is consistent with the transport studies,\cite{Bonn1,Krishana,Hosseini,Puchkov,Takeya}  and in contrast with the earlier ARPES report.\cite{ARPES}  Note that the residual scattering rate in the superconducting state is lower than the normal-state scattering rate extrapolated linearly for $T \to 0$.  In conventional two-dimensional Fermi liquid, the inelastic part of the zero-energy scattering rate decreases for $T \to 0$ as the higher-order power than $T$-linear, $\mathrm{Im}\Sigma_\mathrm{inel}^\mathrm{FL}(0) \propto T^{n}(n>1)$.  Even in marginal Fermi liquid, it decreases linearly,\cite{ARPES,Abrahams}  $\mathrm{Im}\Sigma_\mathrm{inel}^\mathrm{MFL}(0)\propto T$.  So, if the gap opening suppresses only the inelastic scattering, the zero-energy scattering rate should be higher than the linear extrapolation of the normal-state scattering rate.  Therefore, Fig.~\ref{temperature}(b) suggests that the reduction of the scattering rate at $T_c$ occurs not only for the inelastic part but also for the elastic part.  This implies that the elastic impurity scattering is no longer constant as in normal metal, but has a serious energy-dependent effect on the low-energy quasiparticles.  In the normal state, a zero-energy electron is elastically scattered into the other segment of the Fermi surface.   The opening of the $d$-wave superconducting gap closes these scattering channels except for the other nodal points.  Consequently, the nodal quasiparticles are hardly scattered by the impurities in the superconducting state, even though the superconducting gap is closed there.

Then, also the $\omega$-linear behavior of the low-energy scattering rates in Figs.~\ref{enlarged}(d) and \ref{selfenergies}(b) is related with the opening of the $d$-wave gap.\cite{Zhu,Tesanovic,Kastrinakis}  Considering that the electronic density of states is proportional to $|\omega|$ at low energies, the elastic scattering rate may have an $\omega$-linear term,\cite{Zhu} while the inelastic scattering rates only have higher order terms, as convolved with the energy distribution of the coupling modes.  The predominance of the odd scattering process at low energies \cite{Borisenko} is not confirmed by our data.  Although the antibonding Fermi surface is closer to van Hove singularity, the bonding scattering rate is lower than for the antibonding one.  Alternatively, note that the bonding band has more probability amplitude inside the CuO$_2$ bilayer than the antibonding band.  Thus, the bilayer-resolved scattering rates are affected by the spatial distribution of the elastic scatterers: the larger $\omega$-linear term of the antibonding scattering rate indicates that the scatterers are outside the CuO$_2$ bilayer, e.g., antisite defects or excess oxygens in the BiO and SrO layers.\cite{Zhu,Scalapino,Eisaki}  When the impurities are distant from the conduction plane, the nodal quasiparticles are scattered by only small angles into the vicinity of the node, as proposed recently.\cite{Abrahams,Zhu,Scalapino}  Then, with increasing $|\omega|$, the scattering channels are opened linearly, depending on the distance from impurities.\cite{Zhu}  Such small-angle scatterings hardly affect the transport properties, but are important in the single-particle scattering rate.\cite{Durst&Lee}  The present result shows the experimental evidence of the predominance of the forward scattering at low energies.

In conclusion, the ARPES using low-energy tunable photons has unmasked the intrinsic scattering rate and the unique behaviors of the nodal quasiparticles, affected by the elastic scatterings.  The bilayer-resolved quasiparticle properties may provide an internal reference for the mechanics of the quasiparticles in the cuprates.

We thank P.~ J.~Hirschfeld, D.~J.~Scalapino, A.~Kimura and K.~Shimada for enlightening discussion.  This study was performed under the Cooperation Research Program of HiSOR, Hiroshima Synchrotron Radiation Center, Hiroshima University (Proposal No.~04-A-19), and partially supported by Grant-in-Aids for Young Scientists and for Scientific Research in Priority Area ``Invention of Anomalous Quantum Materials'' from the Ministry of Education, Culture, Sports, Science and Technology of Japan.


\begin{thebibliography}{99}
\bibitem{kdependence} When the momentum dependence of $\Sigma(\omega,k)$ is much smaller than the energy dependence, $\mathrm{Im}\Sigma(\omega) = -\frac{1}{2}v_{0} \Delta k$.
\end{thebibliography}
\end{document}